\documentclass[aip,rsi,reprint]{revtex4-1}
\usepackage{amssymb}
\usepackage{amsmath}
\usepackage{textcomp}
\usepackage{graphicx} 
\usepackage{epstopdf}
\usepackage{dcolumn}
\usepackage{bm}
\usepackage{soul}
\usepackage{notes2bib}

\usepackage{lineno}
\def\um{\upmu\mbox{m}}

\def\Wcm2{\mbox{W cm}^{-2}}
\def\Wcmum2{\mbox{Wcm}^{-2}\upmu\mbox{m}^{2}}

\def\cm3{\mbox{cm}^{-3}}

\newcommand{\R}[1]{\textcolor{black}{#1}}

\usepackage{hyperref}
\hypersetup{colorlinks=true,linkcolor=red,citecolor=red,urlcolor=red}
\usepackage{upgreek}

\begin{document}
%\preprint{AIP/123-QED}
\title{Non-invasive characterisation of a laser-driven positron beam}
%%\thanks{Footnote to title of article.}

%% AUTHORS ORDERED BY INSTITUTION, SURNAME (EXCEPT SARRI)
\author{A.~Alejo}
\altaffiliation{Presently at: Clarendon Laboratory, Department of Physics, University of Oxford, OX1 3PU UK}

\affiliation{Centre for Plasma Physics, School of Mathematics and Physics, Queen's University Belfast, BT7 1NN UK}

\author{G.M.~Samarin}
\affiliation{Centre for Plasma Physics, School of Mathematics and Physics, Queen's University Belfast, BT7 1NN UK}

\author{R.~Warwick}
\affiliation{Centre for Plasma Physics, School of Mathematics and Physics, Queen's University Belfast, BT7 1NN UK}

\author{C.~McCluskey}
\affiliation{Centre for Plasma Physics, School of Mathematics and Physics, Queen's University Belfast, BT7 1NN UK}

\author{G.~Cantono}
\affiliation{LIDYL, CEA, CNRS, University Paris Saclay, 91191 Gif Sur Yvette cedex France}

\author{T.~Ceccotti}
\affiliation{LIDYL, CEA, CNRS, University Paris Saclay, 91191 Gif Sur Yvette cedex France}

\author{S.~Dobosz Dufr\'enoy}
\affiliation{LIDYL, CEA, CNRS, University Paris Saclay, 91191 Gif Sur Yvette cedex France}

\author{P. Monot}
\affiliation{LIDYL, CEA, CNRS, University Paris Saclay, 91191 Gif Sur Yvette cedex France}

\author{G.~Sarri}
\email{g.sarri@qub.ac.uk} 
\affiliation{Centre for Plasma Physics, School of Mathematics and Physics, Queen's University Belfast, BT7 1NN UK}

\date{\today}

\begin{abstract}
We report on an indirect and non-invasive method to simultaneously characterise the energy-dependent emittance and source size of ultra-relativistic positron beams generated during the propagation of a laser-wakefield accelerated electron beam through a high-Z converter target. The strong correlation of the geometrical emittance of the positrons with that of the scattered electrons allows the former to be inferred, with high accuracy, from monitoring the latter.
The technique has been tested in a proof-of-principle experiment where, for 100 MeV positrons, we \R{infer} geometrical emittances and source sizes of the order of $\epsilon_{e^+} \approx$ 3 $\mu$m and $D_{e^+} \approx$ 150 $\mu$m, respectively. This \R{is consistent with}  the numerically predicted possibility of achieving sub-$\mu$m geometrical emittances and micron-scale source sizes at the GeV level.

\end{abstract}

\pacs {} % 

\maketitle

\section{Introduction}

In the past decade, significant experimental effort has been put in generating relativistic positron beams using high-power lasers \R{in an all-optical configuration} \cite{alejo2019laser}. Broadly speaking, two main schemes have been adopted \R{in this case}: a direct one, where the laser is directly focussed onto a high-Z thick solid target \cite{Chen1, Chen3, Liang}, and an indirect one, where the laser first accelerates a population of ultra-relativistic electrons via laser-wakefield acceleration (LWFA), which then interact with a high-Z solid target \cite{SarriPRL,SarriNComm,SarriPPCF1,SarriPPCF2,Hafz, Xu2016}. The latter approach has been numerically shown to be able to produce GeV-scale positron energies, with appealing spatial properties \cite{alejo2019}.

The search for novel methods to generate high-energy positrons is mainly motivated by the current need to explore alternative particle acceleration schemes. Currently, the largest particle collider that is operational is the 27 km Large Hadron Collider (LHC) at CERN, which provides proton-proton collisions with a maximum centre-of-mass energy of 13 TeV \cite{CERN}. Before that, the Large Electron-Positron collider (LEP), provided electron-positron collisions with a maximum centre-of-mass energy of 209 GeV \cite{LEP}. Despite several iconic results, including the recent detection of the Higgs Boson \cite{higgs}, there still are several unsolved issues that demand for a higher centre-of-mass lepton collider, ideally in the range of, if not beyond, a TeV.

Several international projects based on radio-frequency technology have been proposed, such as the Compact LInear Collider (CLIC), which is aiming at reaching TeV energies over a \R{13}  km accelerator length \cite{CLIC}. However, the sheer scale of these accelerators is currently imposed by the maximum accelerating field that they can sustain, usually of the order of 10s of MV/m. This makes their realisation considerably expensive and alternative acceleration methods are currently actively studied. Plasma-driven acceleration is arguably one of the most promising schemes, since it can allow for much higher accelerating gradients compared to radio-frequency systems. Landmark results have already been obtained in this area, including 
accelerating fields exceeding 100 GV/m \cite{liu2011all}, the demonstration of energy doubling of a 42 GeV electron beam in less than one meter of plasma \cite{Blumenfeld}, a 2 GeV energy gain of a positron beam in one metre of plasma \cite{Corde}, \R{acceleration in a proton-driven wakefield~\cite{adli2018acceleration}, highly-efficient electron acceleration in a laser-driven wakefield~\cite{litos2014high}, charge-coupling in a multi-stage accelerator~\cite{steinke2016multistage},} and the laser-driven acceleration of electrons up to 8 GeV in only 20 cm of plasma \cite{Gonsalves}.

Large-scale international projects are thus now  studying the feasibility of building a plasma-based electron-positron collider. For instance, plasma-based particle acceleration for the next generation of colliders is included as a major area of investment in the Advanced Accelerator Development Strategy Report in the USA \cite{USroadmap}, it is the main driver for the European consortium ALEGRO \cite{ALEGRO}, and it is one of the main areas of development identified by the Plasma Wakefield Acceleration Steering Committee (PWASC) in the UK \cite{PWASC}.

While the plasma-based acceleration of electrons is rapidly progressing, \R{positron acceleration is far more difficult due to a much narrower region in the wake field suitable for positron acceleration and focusing.} There are four main regimes that are currently being investigated: the quasi-linear regime \cite{blue2003plasma, muggli2008halo}, nonlinear regime \cite{corde2015multi}, hollow channel regime \cite{gessner2016demonstration}, and wake-inversion regime \cite{vieira2014nonlinear,jain2015positron}. 

Whilst each regime has its unique advantages and attractive characteristics, any one of them presents significant challenges that must be overcome before reaching maturity, justifying the considerable attention received by the international research community. One of the major experimental challenges is to provide a positron beam with sufficient spectral and spatial quality, which can then be synchronized with the positron-accelerating region of a plasma wakefield. In particular, one would need low-emittance and short ( $\leq$ 10s of fs) beams with a non-negligible charge ($\geq$ 1 pC). 

It has been recently shown numerically that appealing positron beam characteristics can be achieved by firing a high-energy wakefield-accelerated electron beam through a cm-scale high-Z solid target \cite{alejo2019}. For instance, a 5 GeV, 100 pC electron beam, interacting with a 1cm thick lead target, can produce up to 1 pC of 1 GeV positrons in a 5\% bandwidth, with sub-micron geometrical emittance and a duration comparable to that of the primary electron beam (as short as a few fs \cite{Lundh}). Generating GeV-scale, $\mu$m-size positron beams with sufficiently good emittance would provide experimentalists with an ideal platform to study plasma-based acceleration of positrons; for example, a dedicated experimental area for this kind of work has been included in the Conceptual Design Report for the European plasma-based accelerator facility EuPRAXIA \cite{EuPRAXIA}.

For these studies, it would be highly beneficial to have an online monitoring system for the laser-driven positron beam, where energy, emittance, and source size can be measured on a shot-to-shot basis without interfering with the positron beam. In this laser-driven scheme, the positrons arise from the quantum electrodynamic cascade initiated by the laser-wakefield accelerated electron beam inside the solid target \cite{SarriPRL,SarriNComm,SarriPPCF1}.  The main by-products of this process are also a dense population of gamma-ray photons, and a broadband population of electrons. For high-quality laser-wakefield accelerated electron beams, we show here that the spatial characteristics of the electrons and positrons escaping the solid target are tightly linked. We then propose here to characterise the scattered electrons as a means to infer the positron beam properties in a non-invasive manner. 

The paper is structured as follows: numerical simulations showing the correlation between the properties of the electrons and positrons escaping the converter target are shown in Sec.~\ref{sec:model}. A proof-of-principle experiment will then be discussed, with the experimental setup and the characterisation of the parent electron beam and secondary positrons presented in Sec.~\ref{sec:setup}. The characterisation of the emittance and source size of the electron beam post-converter and how those relate to those of the positrons are discussed in Sec. ~\ref{sec:emittance}. Conclusive remarks are given in  Sec.~\ref{sec:conclusions}.

\section{NUMERICAL MODELLING}\label{sec:model}

\begin{figure}[b!]
	\centering
	\includegraphics[width=\linewidth]{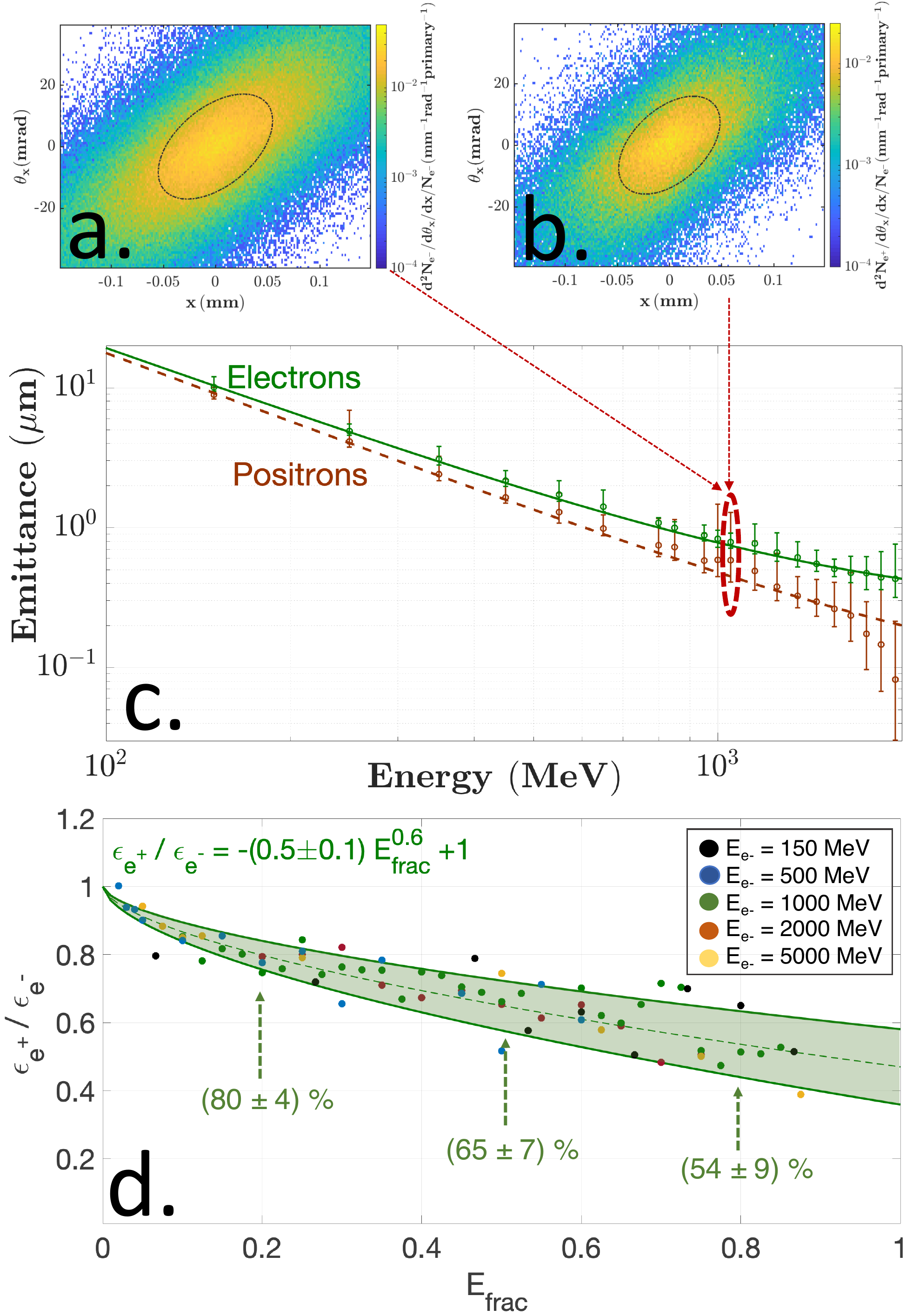}
	\caption{\textbf{a.-b.} Simulated phase-space of 1 GeV electrons (a.) and positrons (b.) at the rear surface of a 10mm Pb converter target irradiated by a 2 GeV mono-energetic electron beam. \textbf{c.} Electron (green) and positron (brown) energy-dependent geometrical emittance at the rear surface of a 10mm Pb converter target irradiated by a 2 GeV mono-energetic electron beam. The dashed red ellipse highlights the values obtained from frames a. and b. \R{Error bars mainly arise from statistical uncertainty in the simulation} \textbf{d.} Ratio between the geometrical emittance \R{of the electrons and positrons escaping the solid target} as a function of their energy normalised by the energy of the primary electron beam (0.15, 0.5, 1, 2, and 5 GeV).}
	\label{fig:model}
\end{figure}

To study the correlation between the electrons and positrons emittance at the rear of the converter target, a series of Monte-Carlo simulations using the scattering code FLUKA \cite{FLUKA1, FLUKA2} have been performed. We simulate $10^7$ mono-energetic electrons contained in a pencil-like beam with different energies: 0.15, 0.5, 1, 2, and 5 GeV. These interact with a 10mm-thick Pb converter (corresponding to approximately 1.8 radiation lengths), where the target thickness has been chosen so to maximise the positron yield at the rear surface \cite{SarriNComm}. In principle, the divergence and source size of the primary electron beam should be included, since they might affect the spatial properties of the particles escaping the converter target. \R{However, as shown later, the geometrical emittance of the positrons escaping the target is of the order of a few microns, as dictated by the spread induced by the quantum electrodynamic cascade inside the converter. As long as the emittance of the primary electron beam is much smaller than this value, as usually is the case in laser-wakefield acceleration \cite{Brunetti}, it can be ignored.}
\R{ As an example, we show, in the supplementary material \bibnote{\R{\emph{--url to be inserted here in published version--.} Simulated ratio between the emittance of the positrons and electrons escaping the converter target as a function of energy for a primary electron beam with zero (solid line) and a 5 mrad divergence (empty circles). $E_{frac}$ denotes the ratio between the particle energy and that of the primary electrons}}, a negligible difference between the calculated positron emittance for a primary electron beam with a 5 mrad divergence and that for a primary electron beam with zero initial divergence.}
%\st{This value is much smaller than the emittance of the positrons and the scattered electrons after the solid target, and can thus be safely neglected.}

Nonetheless, it is still to be intended that the results shown here are only for demonstration purposes and will be used to infer the positron emittance and source size for our proof-of-principle experiment. \R{Even though the same qualitative behaviour will hold, slightly quantitative differences in the results} will be obtained for each specific setup to be adopted (e.g., different converters and different spectra of the parent electron beam) and numerical modelling of the specific configuration to be used should be performed before implementing this technique.

An example of the simulation results is shown in Fig. \ref{fig:model}. The electron and positron geometrical emittances (examples in frames \ref{fig:model}.a and \ref{fig:model}.b) are strongly energy-dependent following a decreasing power law, in agreement with recently published numerical results \cite{alejo2019}. Interestingly, the positron emittance is seen to be consistently smaller than that of the scattered electrons (see, for example, Fig. \ref{fig:model}.c). 

This can be intuitively understood with the following reasoning. \R{For a target thickness ($L_c$) of the order of a radiation length ($L_{RAD}$), positrons in the target are mainly generated following a two-step process (bremsstrahlung + pair production in the nuclear field) whereas the scattered electrons can be either generated during pair production or during the production of bremsstrahlung radiation. However for these target thicknesses the number of electrons generated by pair production can be ignored, resulting in the population of electrons escaping the solid target arising almost exclusively from scattering of the primary electron beam. On average, the positrons are thus created deeper into the target and exit, for each defined energy, with a smaller source size.}

\R{These assumptions break down in the limiting cases of ultra-thin or thick targets. As previously discussed \cite{SarriPRL,SarriPPCF1}, for $L_c/L_{RAD}\leq10^{-2}$ , direct electro-production (sometimes referred to as the trident process) \cite{bhabha1935creation} will dominate resulting in pairs being generated directly as an electron traverses the nuclear field, without the intermediate step of generating a real photon via bremsstrahlung. On the other hand, thick targets will allow for multi-step cascades up to a point where the number of escaping electrons and positrons will become approximately equal, since they both arise from pair production. In this case, occurring at approximately $L_c/L_{RAD}\geq 5 $, \cite{SarriNComm} the emittance of the escaping electrons and positrons will be approximately equal. The positron emittance will thus be smaller than that of the scattered electrons as long as we can neglect trident pair production and multi-step cascades, i.e., for $ 10^{-2} \leq L_c/L_{RAD} \leq 5$. In the case of lead, this corresponds to  $ 60 \mu$m $\leq L_c \leq 2.5$ cm.}

%For a target thickness of the order of 2 radiation lengths, direct electroproduction (sometimes referred to as the trident process) \cite{bhabha1935creation} can be ignored, since it is dominant only for target thicknesses $d\leq 10^{-2}$ $L_{Rad}$ \cite{SarriPRL,SarriPPCF1} and multi-step cascades are unlikely. Positrons are thus generated following a two-step process (bremsstrahlung + pair production in the nuclear field) whereas the scattered electrons can be either generated during pair production or during the production of bremsstrahlung radiation. However for this target thicknesses the number of electrons generated by pair production can be ignored, resulting in the population of electrons escaping the solid target arising almost exclusively from scattering of the parent electron beam. \R{This is valid as long as multi-step cascades are unlikely, as expected for a target thickness of the order of 1-2 radiation lengths. For thicker targets, i.e. $d > 5 L_{rad}$ this will cease to be valid and approximately equal numbers of electrons and positrons will escape the converter, both arising from pair production. In this case the two emittances will be practically identical.}

%On average, the positrons are thus created deeper into the target and thus exit, for each defined energy, with a smaller source size. This is shown in 
Figure \ref{fig:model}.d depicts the ratio between the positron and electron geometrical emittance as a function of their energy. The results are seen to be fairly independent of the initial energy of the primary electron beam, but are mainly related to the ratio of the energy of the escaping particle with that of the primary electrons.
This gives, in the ultra-relativistic regime, a general scaling between the electron and positron emittance, practically independent from the energy of the parent electron beam. The trend extracted from the simulations is in the form of a power law: $\epsilon_{e^+}/\epsilon_{e^-} = - (0.5\pm0.1)E_{frac}^{0.6} +1$, with $E_{frac}$ the ratio between the particle energy and that of the parent electrons. A certain level of uncertainty is present, mostly due to the non-ideal statistics in extracting the positron emittance from the simulations \R{as illustrated by the error bars in Fig. \ref{fig:model}.c.} 

\section{Experimental setup}\label{sec:setup}

\begin{figure}[b!]
	\centering
	\includegraphics[width=\linewidth]{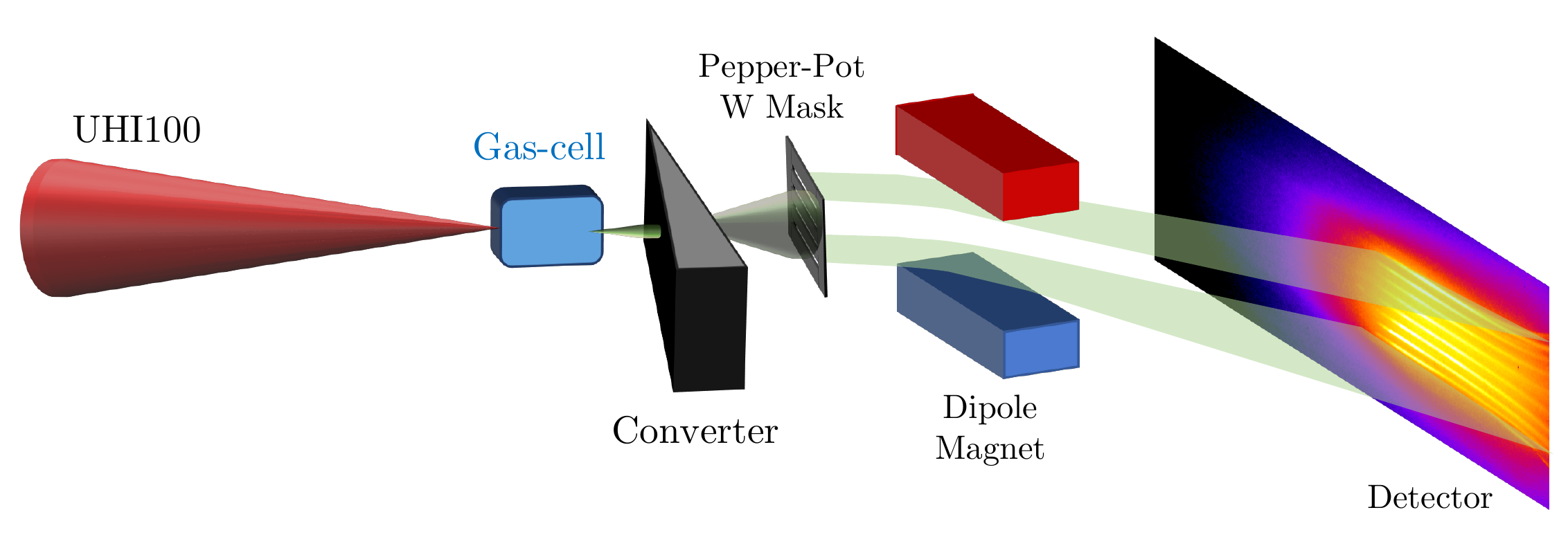}
	\caption{\textbf{Experimental setup.} Cartoon depicting the experimental setup. The UHI100 laser was focussed using an F/13 OAP onto a gas cell of variable length. The electron beam driven by the laser was the impinged into a secondary Pb converter of variable thickness. The electron-positron beam was then propagated through a pepper-pot mask before entering a magnetic spectrometer.}
	\label{fig:setup}
\end{figure}

In order to experimentally check the viability of this technique, an experiment was carried out using the UHI-100 laser facility at CEA Saclay. The system delivers laser pulses with an energy of $E_p=2.5\pm0.1\,$J before compression ($\sim0.9\,$J in focus), a duration of $\tau_p= 24\pm2\,$fs, and a central wavelength of $\lambda_0=800\,$nm. The  laser was focussed down to a $28\,\um$ FWHM (Full Width at Half Maximum) spot using an $F/15$ off-axis parabolic mirror combined with an adaptive optic, producing a focal spot with a peak intensity $I_0\simeq(4.7\pm0.7)\times10^{18}\,\Wcm2$ (dimensionless amplitude $a_0 = 1.5\pm0.1$). The laser was focussed onto the entrance of a gas-cell, with a variable length ranging from 0 to 5 mm, and filled with a H$_2$+$5\%$N$_2$ gas mix. The experimental setup is schematically depicted in Fig.~\ref{fig:setup}.  

 The laser-driven electron beam was then made to propagate through a wedge-shaped Pb converter, placed 13\,mm downstream from the rear of the gas cell. The thickness of the converter could be varied in the range 1-20 mm by laterally displacing the wedge.  Both the spatial and spectral properties of the leptonic beam were characterised. The spectral characterisation was performed by using a magnetic spectrometer, consisting of a $50\,$mm, $0.8\,$T dipole magnet and detector screen. The magnet was placed at a distance of $124\,$mm downstream of the back of the gas cell, whereas the detector screen was placed $384\,$mm away from the gas cell. Image plates (Fujifilm BAS-MD) were used as detectors for signal accumulation, whereas Lanex screens were used for single-shot measurements. %The lanex screens were imaged using an 12-bit EMCCD camera (Andor iStar) and zoom lens. 
 Suitable shielding (not shown) was inserted to minimise the background noise at the detectors. In addition to this, the magnet could be removed from the beamline to characterise the spatial properties of the electron beam, as well as its pointing fluctuations. To do so, an additional Lanex screen was placed at a distance of 1083\,mm from the target (not shown).

The emittance of the beam was characterised by using the pepper-pot technique \cite{pepperpot}. In particular, due to the strong dependence of the emittance on the particle energy, a 1D pepper-pot was employed. In this case, an array of slits is used to select a number of beamlets, which are then propagated through the spectrometer, allowing to retrieve an energy-resolved measurement of the emittance. \R{It must be noted that the pepper-pot technique is known to overestimate the emittance of a beam in which the position and momentum of the particles are strongly correlated \cite{downer2018diagnostics}, as could be the case in a laser-wakefield accelerator. Even though this correlation is expected to rapidly degrade during the propagation in the converter target, the values reported here should still be considered as an upper limit for the geometrical emittances.}

%\R{Due to experimental constraints, the pepper-pot measurement could only performed along one direction

\begin{figure}[t!]
	\centering
	\includegraphics[width=\linewidth]{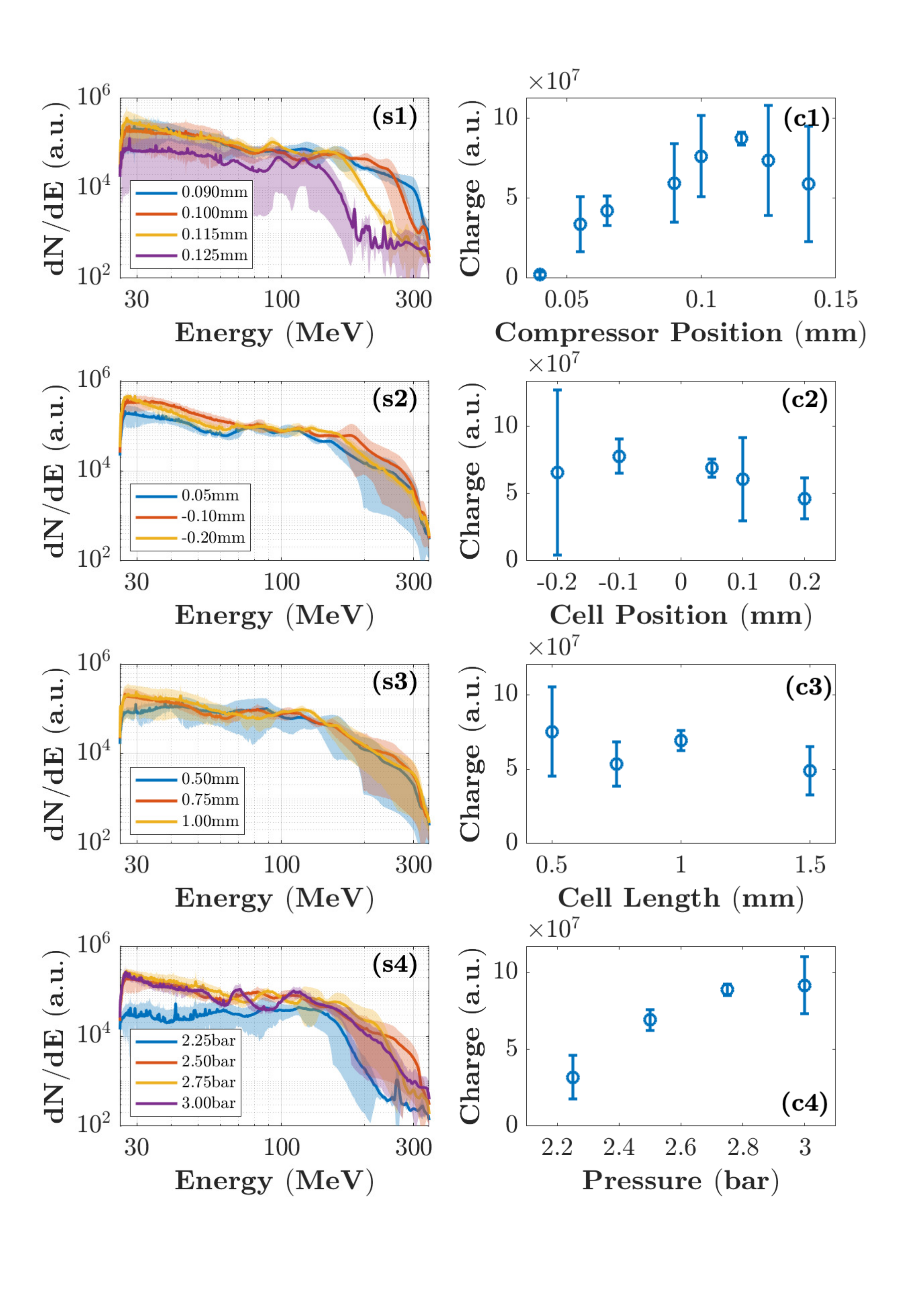}
	\caption{\textbf{Optimisation of the primary electron beam}. Average spectra (left) and charge (right) of the electron beams produced when different parameters of the laser-plasma interaction were modified, namely (1) the compressor position (shortest pulse corresponds to 0\,mm), (2) the position of the gas cell entrance with respect to the laser focal plane, (3) the length of the gas cell, and (4) the backing pressure of the gas filling the cell. The shaded areas in the spectra plots, and the error bars in the charge plots, represent the standard deviation of the data with respect to the average.}
	\label{fig:electron}
\end{figure}

The spectral properties of the leptonic beam after the converter has been shown to be fairly independent of the spectral shape of the parent electron beam~\cite{alejo2019}. For this reason, the laser-plasma interaction was optimised to produce primary electron beams with the highest possible charges and energies in an ionisation-injection regime \cite{clayton2010self}. This was achieved for a gas-cell backing pressure of 2.75 bar (corresponding to an electron density of $n_e\approx 10^{19}$ cm$^{-3}$), gas-cell length of 1mm, and a detuning of the optimum  compressor grating position %(i.e., the position that guarantees the minimum laser pulse duration) 
of 0.1 mm. A summary of the parameter scan for the parent electron beam is shown in Fig. \ref{fig:electron}, where the left column (Figs.~\ref{fig:electron}(s1-s4)) depicts the average spectrum for each set of parameters, with the shaded area representing the standard deviation of the data from the average. The right column (Figs.~\ref{fig:electron}(c1-c4)) shows the dependence of the total electron charge in the beam on each of the parameters. 

%Since the minimum detectable energy that could be measured by the spectrometer was $\sim50\,$MeV, the total charge was obtained by integrating the signal on the pointing monitor, when the dipole magnet was moved away from the beam axis, allowing to account for the charge in the full spectrum.

In these experimental conditions, a reproducible electron beam with the following characteristics was then obtained and used for the rest of the experiment: maximum energy $E_s = (200\pm20)$ MeV, divergence $\theta_s = (5\pm1)$ mrad\bibnote{\R{\emph{--url to be inserted here--.} Measured energy-dependence of the primary electron beam}}, and a total charge of the order of 10s of pC. Assuming that we were operating in a heavily loaded blow-out regime, the electron beam duration can be estimated to be: $\tau_s \approx 2\sqrt{a_0}/\omega_p\lesssim 12$ fs, with $\omega_p= 2\times10^{14}$ Hz the plasma frequency of the background gas. Similarly, \R{the upper limit for the electron beam source size is given by the size of the accelerating bubble in the wakefield:} $D_s \lesssim 2c\sqrt{a_0}/\omega_p\simeq 4$ $\mu$m \cite{Thomas}. \R{Even though compressor de-tuning might induce cilindrically asymmetric electron beams \cite{Popp}, these deviations appear negligible in our experimental setup \bibnote{\R{\emph{--url to be inserted in  the published version--.} Typical transverse spatial distribution of the undeflected primary electron beam from the wakefield accelerator, showing a symmetrical distribution with a divergence of 5 mrad. A population of electrons on the left-hand side can be seen but this is due to the influence of the dipole magnet, which, due to the limited size of the vacuum chamber, could not be moved sufficiently away from the electron propagation axis, and thus could still have a sizeable effect on low energy electrons.}}, justifying the cylindrical symmetry assumed throughout this work.}  

%\section{Positron generation}\label{sec:positron}

\begin{figure}[t!]
	\centering
	\includegraphics[width=\linewidth]{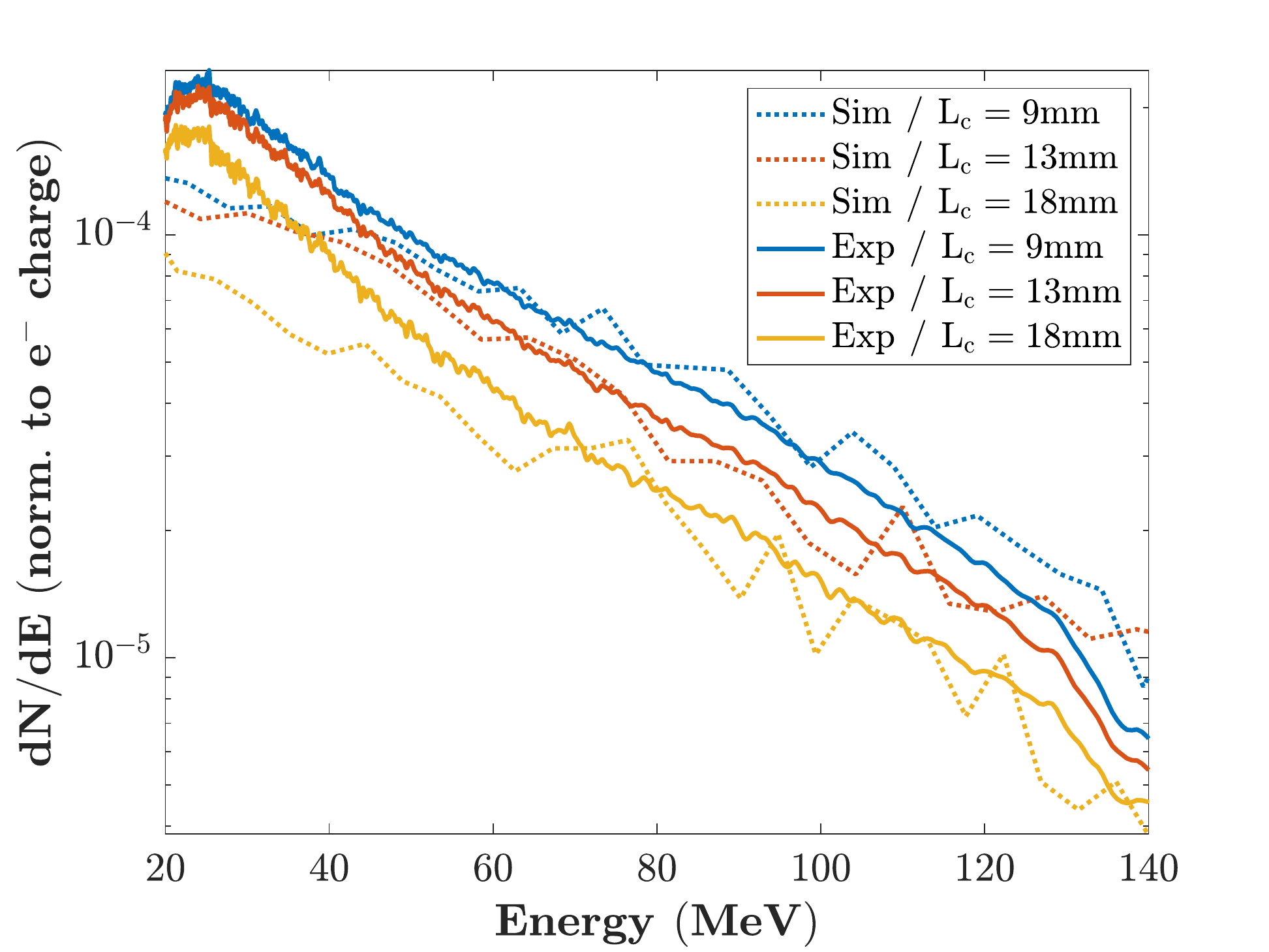}
	\caption{\textbf{Positron spectra}. Average spectra of the positron beam generated after the propagation of an electron beam through a Pb converter thickness \R{$L_c$ of} 9, 13 and 16\,mm, respectively. Each spectrum is obtained as the average of 10 consecutive shots, and the total signal is normalised to the signal measured for the parent electron beam.}
	\label{fig:positron}
\end{figure}

The electron beam was then directed onto a Pb converter of variable thickness. The resulting positron beam was characterised by using the same magnetic spectrometer as in Sec.~\ref{sec:setup}. The experimentally measured positron spectra for different converter thicknesses are shown in Fig. \ref{fig:positron}, compared with those resulting from FLUKA simulations \cite{FLUKA1, FLUKA2}, showing good quantitative agreement.

The spectra resemble a relativistic Maxwellian distribution, with a maximum positron charge observed for a converter thickness of $\sim9\,$mm, which corresponds to approximately 1.6 radiation lengths ($L_{Pb}\simeq 5.6\,$mm). This observation is in good agreement with previous experimental results \cite{SarriPRL,SarriNComm,SarriPPCF1,Hafz, Xu2016}. Numerical simulations indicate a temporal broadening of the positrons at 100 MeV of the order of 50 - 100 fs, resulting in a peak positron current, at the rear surface of the converter, of the order of 1 A. %While the low charge contained in the positron beam ($\approx$) made it hard to directly measure its emittance and source size, these quantities are inferred from those of the scattered electrons, as discussed in Section VI. 

\section{Emittance and source size after the converter}\label{sec:emittance}

For our experimental parameters, the quantum cascade process has a relatively low efficiency, resulting in a sub-pC positron charge. This low charge, in addition with the high-noise environment, prevents from directly characterising the emittance of the positron beam using a pepper-pot mask. For this reason, since the emittance of the electron and positron beams generated during the cascade are strongly correlated, the emittance of the electron beam after the converter was characterised instead.

\begin{figure}[t!]
	\centering
	\includegraphics[width=\linewidth]{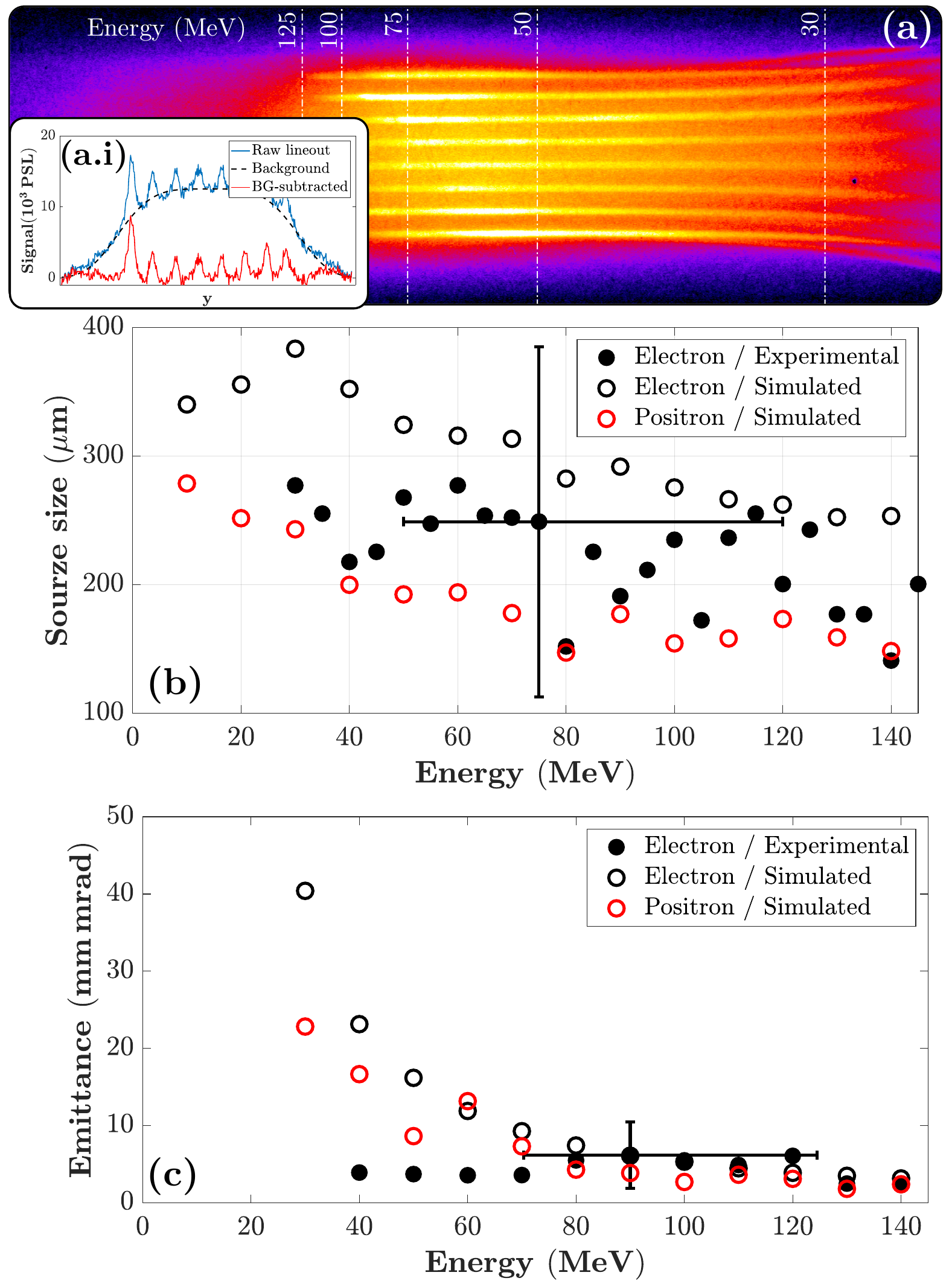}
	\caption{\textbf{Electron emittance and source size}. \textbf{(a)} Typical raw image of the electron signal after propagating through the pepper-pot mask and the dipole magnet. \R{Inset \textbf{(a.i)} depicts the lineout corresponding to $E=50\,$MeV, where the blue line is the raw lineout, the dashed black line shows the calculated background, and the red line shows the signal after background subtraction, which was used to calculate the source size and emittance at this energy.} \textbf{(b)} Extracted energy-dependent source size of the electron beam, compared to the simulated electron and positron source sizes. \textbf{(c)} Extracted energy-dependent geometrical emittance of the electron beam compared with the simulated electron and positron geometrical emittances. For clarity, exemplary error bars are shown for a single experimental data point in (b) and (c).}
	\label{fig:emittance}
\end{figure}

For the emittance measurement, a pepper pot mask was placed along the electron beam propagation axis. 
A typical raw image of the scattered electrons after propagation through the mask and dispersion by the magnetic spectrometer is shown in Fig.~\ref{fig:emittance}(a). Each of the horizontal lines visible in the figure corresponds to a beamlet propagating through a single aperture in the pepper pot mask, which is spectrally resolved along the horizontal direction inside the spectrometer.

The electron source size was estimated by means of a penumbral imaging technique, in which spatial information is recovered from the shadow produced by an aperture (\R{see inset in Fig. \ref{fig:emittance}.a}), such as the slits in our case. %In particular, the extent and profile of the shadow are given by a convolution of the source size, the system magnification, slit function, and detector resolution. 
Considering that the spatial profile of the source can be well approximated by a Gaussian, the FWHM of the source is given by the distance between the points at which the signal is $12\%$ and $88\%$ of the maximum signal, respectively. The source size can be thus estimated by measuring such distance at the detector, and transforming to the source plane by taking into account the magnification, slit function, and detector resolution.

The dependence of the electron source size with the electron energy is shown in Fig.~\ref{fig:emittance}(b), alongside a comparison with the electron and positron source sizes obtained using FLUKA simulations \R{assuming the experimentally-measured primary electron beam}. The uncertainties for the source size consider the slit size, resolution of the detector, and magnification of the system; whereas the main contribution towards the uncertainty in energy is given by the size of the slits on the pepper pot mask, leading to a possible overlap of an energy range at a given point on the detector. The source size is found to slowly vary with the particle energy with an average value of $(230\pm100)\um$.  As it can be seen in Fig.~\ref{fig:emittance}(b), both the measured source size and its quasi-constant value in the energy range of interest are in good agreement with the expected values obtained in the simulations. It is of particular interest to note that the positron source size $D_p$ at a defined energy is consistently lower than that of the electrons $D_e$ in the energy range between 10 and 140 MeV ($D_p/D_e = 0.64\pm0.07$). In this range, the positron source size slowly diminishes with positron energy, from a value of 280 $\mu$m at 10 MeV down to 150 $\mu$m at 140 MeV. These values are consistent with recently reported numerical modelling \cite{alejo2019}. 

The emittance of the scattered electron beam was extracted using the pepper-pot equations~\cite{pepperpot}, taking into account that a 1-D system was used (\R{see Fig. \ref{fig:emittance}.c}). Similarly to the case of the source size, the emittance is found to be approximately constant for the energy range considered, with an average value of $\epsilon_{xe}\sim 3.5\,$ $\mu$m (normalised emittance at 100 MeV of $\epsilon_{ne}\approx 200$ $\pi$ $\mu$m). The measured values are in good agreement with FLUKA simulations, considering the relatively large uncertainty for the energy of the particles at a given position on the detector, particularly in the case of lower energy particles, for which magnetic fringe fields could introduce additional sources of error in the measurement.

Again, the positron geometrical emittance appears consistently lower than that of the scattered electrons. For example, at 80 MeV ($E_{frac} = 0.4$) the ratio between the simulated positron emittance and the measured emittance of the scattered electrons is 66\%, in good agreement with the $(69\pm6)$\% predicted by the trend shown in Fig. \ref{fig:model}. 

\section{Conclusions}\label{sec:conclusions}
We report on a non-invasive method to characterise the positron beam generated during the interaction of a laser-wakefield electron beam through a high-Z converter. In the ultra-relativistic regime, \R{and for converter thicknesses of the order of a radiation length}, the positron geometrical emittance is found to be consistently smaller than that of the scattered electrons, with general trend that is virtually independent from the energy of the primary electron beam. For a $\approx$ 10 pC broadband electron beam with a maximum energy of  200 MeV, the positron beam is found to exit the converter target with a sub-pC charge, a broadband spectrum extending up to 140 MeV, a duration of the order of 100 fs, and a geometrical emittance at 100 MeV of $\epsilon_{e^+}\sim 3.5\,$ $\mu$m. 
The results \R{confirm} numerical work recently published, \R{and are thus consistent with the potential of state-of-the-art PW-scale laser systems to generate GeV-scale positron beams with fs duration and sub-micron geometrical emittance from a fully optical configuration}. Positron beams of similar characteristics, once energy-filtered by means of conventional magnetic elements, will be usable as test beams to study advanced plasma-based positron acceleration schemes.

\section*{Acknowledgements}

The authors acknowledge financial support from EPSRC (Grant No: EP/N027175/1 and EP/P010059/1).
This project has received funding from the European Union$’$s Horizon 2020 Research and Innovation programme under Grant Agreement No 730871.
Authors would also like to acknowledge the support of the mechanical engineering and laser staff of the CEA Laboratory (Saclay, France).

\bibliography{bibliography}
\end{document}